\documentclass[12pt]{iopart}

\usepackage[dvipdfmx]{color}
\usepackage[dvipdfmx]{graphicx}
\usepackage{bm}
\usepackage{times} 
\usepackage{cite}

\usepackage[mathlines, columnwise, pagewise]{lineno}

\usepackage[utf8]{inputenc} 
\usepackage[T1]{fontenc} 
\usepackage{mathptmx} 

\begin{document}

\title[]
{\textcolor{black}{Magnetic entropy change of ErAl$_{2}$ magnetocaloric wires
fabricated by a powder-in-tube method}}

\author{Takafumi D. Yamamoto$^{1}$, Hiroyuki Takeya$^{1}$, Suguru Iwasaki$^{1}$,
Kensei Terashima$^{1}$, Pedro Baptista de Castro$^{1, 2}$, Takenori Numazawa$^{1}$, and Yoshihiko Takano$^{1, 2}$}

\address{$^{1}$National Institute of Materials Science, Tsukuba, Ibaraki 305-0047, Japan}
\address{$^{2}$University of Tsukuba, Tsukuba, Ibaraki 305-8577, Japan}

\ead{YAMAMOTO.Takafumi@nims.go.jp}

\begin{abstract}
\textcolor{black}{
We report the fabrication of ErAl$_{2}$ magnetocaloric wires
by a powder-in-tube method (PIT) and the evaluation of magnetic entropy change
through magnetization measurements.
The magnetic entropy change of ErAl$_{2}$ PIT wires exhibits similar behavior
to the bulk counterpart, while its magnitude is reduced by
the decrease in the volume fraction of ErAl$_{2}$
due to the surrounding non-magnetic sheaths.
We find that another effect reduces the magnetic entropy change of
the ErAl$_{2}$ PIT wires around the Curie temperature,
and discuss its possible origin in terms of a correlation between
magnetic properties of ErAl$_{2}$ and mechanical properties of sheath material.
}
\end{abstract}


\submitto{\JPD}

\maketitle

\section{Introduction}
Magnetic refrigeration is a cooling technology
based on the magnetocaloric effect
in which the variation in magnetic entropy (or temperature) of a magnetic material is
caused by changing a magnetic field.
A well-established applied technique is a cooling by adiabatic demagnetization
to achieve ultra-low temperatures below 1 K\cite{Debye-AnnPhys-1926,Giauque-JAC-1927}.
Since 1997, the application to room temperature refrigerators has been enthusiastically studied
because magnetic refrigeration has the potential to outperform the conventional vapor-compression refrigeration
concerning energy efficiency and environmental friendliness\cite{Zimm-ACE-1998, Bruck-JPD-2005}.
Many great efforts have been made up to date on the development of
working materials with a large magnetocaloric effect near room temperature
\cite{Gschneidner-RPP-2005, Tishin-JMMM-2007, Franco-PMS-2018}
(\textit{e.g.}, Gd$_{5}$Si$_{2}$Ge$_{2}$ discovered by Pecharsky and Gschneidner\cite{Pecharsky-PRL-1997})
and efficient refrigeration systems such as an active magnetic regenerator (AMR)
\cite{Barclay-Patent-1982, Gschneidner-IJR-2008,Nielsen-IJR-2011}.

A newly attracting potential application of magnetic refrigeration is the hydrogen liquefaction.
Hydrogen is one of the cleanest energy sources to replace fossil fuels\cite{Jones-Science-1971}.
For the use in society, it is efficient and economical
to transport and store hydrogen in a liquid state
because liquid hydrogen is denser than gaseous hydrogen.
In this context, high-efficient liquefaction technology is required.
One of the authors has confirmed  >50\% liquefaction efficiency
in a test apparatus of the Carnot magnetic refrigerator
worked around the hydrogen liquefaction temperature (20.3 K)\cite{Kamiya-Cryoc-2007}.
On the other hand, in the practical liquefaction process,
it is necessary to pre-cool the hydrogen gas
from the temperature of a heat sink, such as liquid nitrogen, to nearly 20.3 K
by using a multistage AMR cycle
\cite{Utaki-Cryoc-2007,Matsumoto-JPCS-2009,Numazawa-Cryo-2014}.
What should be noted here is that the magnetocaloric material must be processed into
a specific shape suitable for each refrigeration system.
For example, spherical particles or thin plates have been employed for AMR systems
to gain better heat exchange efficiency between the working material and the heat-exchanger fluid
\cite{Yu-IJR-2010, Tusek-IJR-2013}.

Candidate materials for hydrogen magnetic refrigeration are often found
in intermetallic compounds containing heavy rare-earth elements.
A representative example is the lanthanide (R) Laves phase RT$_{2}$ (T $=$ Al, Co, and Ni)
\cite{Hashimoto-ACM-1986,Tomokiyo-ACM-1986,Zhu-Cryo-2011},
which exhibits a large magnetic entropy change in the temperature range from 20 to 80 K.
However, these compounds are difficult to be shaped
due to their poor ductility and malleability.
Moreover, these materials are quite brittle, leading to a risk of damage by the friction
between them during the refrigeration cycle operation.
Such mechanical properties prevent these candidate materials from
being used as magnetic refrigerants.
Besides, they are known to easily absorb hydrogen,
resulting in the degradation of the refrigerants and their performance.
A coating for protection is a typical way to solve this issue,
\textcolor{black}{
but this takes extra effort in addition to the shaping process
for producing magnetic refrigerants.}

Very recently, Funk et al. reported\cite{Funk-MTE-2018}
a way for producing magnetocaloric wire by a PIT method in La (Fe, Co, Si)$_{13}$,
which is a promising material for the room temperature magnetic refrigeration
\cite{Hu-APL-2001,Fujieda-APL-2002}.
The PIT method is a conventional and simple technology that has been developed
in the field of superconducting wires\cite{Kunzler-PRL-1961,Kunzler-RMP-1961},
in which a powdered raw material is filled into a metal tube and
then formed into wire-shaped by various metal workings.
This approach is attractive because of many advantages
in applying the PIT method to the candidate materials
for hydrogen magnetic refrigeration as follows:
(1) This method is available even for the difficult-to-process materials
since raw materials can be powder.
(2) The metal sheath surrounding the magnetic refrigerants protects
them from the friction wear or the hydrogen embrittlement.
(3) As Funk et al. have pointed out, the wires provide the possibility of
various arrangements of magnetic refrigerants.
Besides, it should be noted that recent works have focused on wire-shaped magnetocaloric materials
because they have been suggested to show superior performance as magnetic refrigerants
to conventional spherical or plate-like materials\cite{Shen-APL-2016,Ueno-Fujikura-2017,Vuarnoz-ATE-2012}.

In this paper, we investigate the effects of a PIT process on
the magnetocaloric properties in a well-studied compound ErAl$_{2}$
that exhibits a second-order ferromagnetic transition at $T_{\rm c} \sim$ 14 K
\cite{Hashimoto-ACM-1986,Nereson-JAP-1968,Pecharsky-JAP-1999}.
We have confirmed that the magnetic entropy change $\Delta S_{M}$ is similar
in the ErAl$_{2}$ PIT wires and the bulk counterpart,
while it decreases in magnitude for the former due to
a reduction of volume fraction of ErAl$_{2}$ in the wires.
We have further found that another effect causes
an additional decrease of $\Delta S_{M}$ near $T_{\rm c}$,
which depends on the sheath material.
This is the first report to apply the PIT method
for fabricating magnetocaloric wire for the hydrogen liquefaction.

\section{Experimental details}
\begin{figure}[b]
\centering
\includegraphics[width=85.00mm]{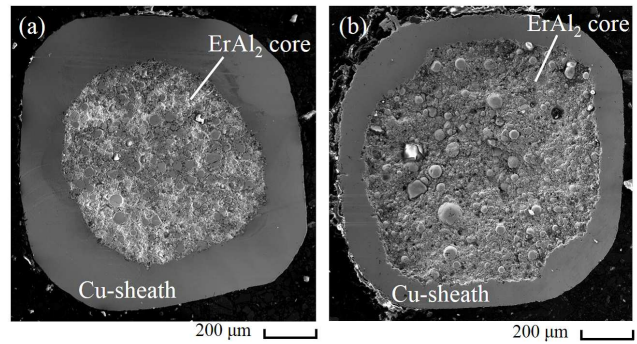}
\caption{Cross-sectional SEM images of ErAl$_{2}$/Cu PIT wires
fabricated from (a) 6$\times$4 and (b) 6$\times$5 tubes (see the text).
The size is about 1 mm each.}
\label{fig:SEM}
\end{figure}

ErAl$_{2}$ single-core wires were fabricated by
an \textit{ex-situ} PIT method without any heat treatment.
ErAl$_{2}$ raw powder with a diameter of less than 50 $\mu$m was
prepared by a gas-atomization process.
The powder was filled into several metal tubes with
50 mm in length, an outer diameter ($d_{\rm o}$) of 6 mm,
and inner diameter ($d_{i}$) of 4 or 5 mm
(hereafter, referred to 6$\times$4 tube and 6$\times$5 tube respectively).
The tubes were plugged on both sides by
cylinders 7 mm in length made of the same material as each tube.
Thus-made initial rods were first groove-rolled
into wires with a size of 2 mm stepwisely.
Then the wires were cut into about 70 mm
and further groove-rolled into those with a size of 1 mm stepwisely.
The resulting PIT wires were 260-300 mm in length.
Cu, Al, and Brass were employed as the sheath materials
because they are non-magnetic and show relatively high thermal conductivity.

The cross-sectional observations for the fabricated PIT wires were carried out
using a JEOL JSM-6010LA scanning electron microscope (SEM) operated at 15 kV.
The cross-sectional area was evaluated using an image analysis software Image-J
(National Institute of Health, US).
Figures \ref{fig:SEM}(a) and \ref{fig:SEM}(b) show SEM images of ErAl$_{2}$/Cu PIT wires
fabricated from the 6$\times$4 and 6$\times$5 tubes, respectively.
These images indicate that the ErAl$_{2}$ powder is uniformly filled
inside the Cu-sheath as a core material.
The cross-section ratios of the ErAl$_{2}$ core to the whole wire were
evaluated to be 0.437 from Fig. \ref{fig:SEM}(a) and 0.655 from Fig. \ref{fig:SEM}(b),
which are comparable to the theoretical filling rate, defined as $d_{\rm i}^2/d_{\rm o}^2$, 
expected for each initial tube (0.444 for the 6$\times$4 tube and 0.694 for the 6$\times$5 tube).
This result implies that the core and the sheath material were deformed
at the same proportion during the rolling process. 
We have found the same features in ErAl$_{2}$/Al and ErAl$_{2}$/Brass PIT wires.

Magnetization measurements were performed by
a Quantum Design magnetic property measurement system.
Temperature ($T$) dependence of magnetization ($M$) of the ErAl$_{2}$ powder
and the PIT wires was measured between 2 and 60 K
at a temperature sweep rate of 0.5 K/min
under various magnetic fields ($\mu_{0}H$)
ranging from 0.1 to 5 T in zero-field cooling (ZFC) process.
The magnetic fields were applied along the longitudinal direction
of each PIT wire with 5-7 mm in length.
For the powder sample, field dependence of magnetization was collected between
0 and 5 T in the temperature range of 2 $\leq T \leq$ 40 K.

The magnetic entropy change is often evaluated
from the isothermal magnetization ($M$-$\mu_{0}H$) measurements
by using one of Maxwell's relations
\begin{equation}
\Delta S_{M}(T, \mu_{0}\Delta H) =
\mu_{0} \int_{H_{i}}^{H_{f}}{\left(\frac{\partial M}{\partial T}\right)_{H}}dH,
\label{eq:deltaS}
\end{equation}
where $H_{i}$ and $H_{f}$ is the initial and final magnetic field,
and $\Delta H = H_{f} - H_{i}$.
However, this way requires us to collect lots of magnetization curves
at various temperatures for correct evaluation,
which is somewhat time-consuming and makes it difficult
to obtain in detail the temperature dependence of $\Delta S_{M}$.
So we first examined how to efficiently and accurately evaluate $\Delta S_{M}$
from the isofield magnetization ($M$-$T$) measurements in the ErAl$_{2}$ powder.
\textcolor{black}{
Then the validity of this unconventional method was verified
by comparing the results obtained from this and the often-used method.}
In the following, $H_{i}$ is set to zero.
\section{Results and discussion}
\begin{figure}[t]
\centering
\includegraphics[width=80.00mm]{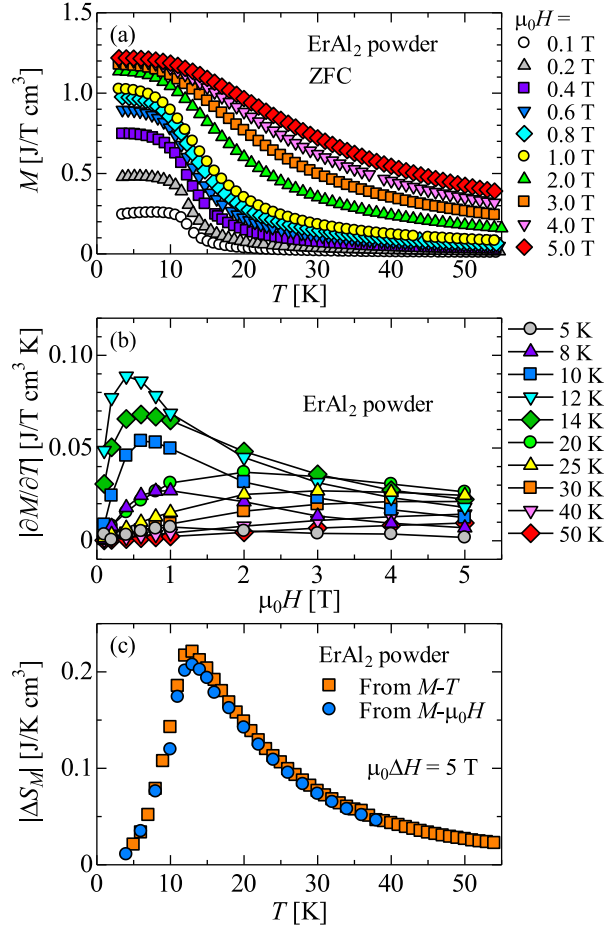}
\caption{(Color Online) (a) Temperature dependence of magnetization
and (b) field dependence of the temperature derivative of
magnetization for the ErAl$_{2}$ powder.
\textcolor{black}{(c) Magnetic entropy change for $\mu_{0} \Delta H =$ 5 T
as a function of temperature evaluated from $M$-$T$ (square)
and $M$-$\mu_{0} H$ (circle) measurements.}}
\label{fig:powder}
\end{figure}

Figure \ref{fig:powder}(a) shows the $M$-$T$ curves of the ErAl$_{2}$ powder.
One finds the features typical of a second-order ferromagnetic transition with $T_{\rm c}$ of 12 K,
defined as the temperature at which $|\partial M/\partial T|$ at 0.1 T takes a maximum.
The slight discrepancy with the $T_{\rm c}$ in the literatures may be
because that the ErAl$_{2}$ powderwas made by the gas-atomization process
in which where the material is quenched. 
Similar $M$-$T$ curves have been observed in all the ErAl$_{2}$ PIT wires (not shown).
To calculate $\Delta S_{M}$ correctly from $M$-$T$ measurements,
one should select the measuring magnetic fields properly.
As shown in Fig. \ref{fig:powder}(b),
$|\partial M/\partial T|$ calculated from Fig. \ref{fig:powder}(a)
exhibit a non-monotonic field dependence, especially around $T_{\rm c}$:
it steeply increases and reaches the highest point below 1 T,
followed by a gradual decrease under higher fields.
Since $\Delta S_{M} (T, \mu_{0} \Delta H)$ at a fixed $T$ is equivalent to
the area in the $\partial M/\partial T$-$\mu_{0} H$ plane,
this peak structure can largely affect the evaluated value of $\Delta S_{M}$.
Accordingly, it is essential to finely collect the $M$-$T$ curves under magnetic fields
in which the peak of $|\partial M/\partial T|$ appears\cite{Suppl-info}.
\textcolor{black}{Figure \ref{fig:powder}(c) shows $\Delta S_{M}$($T$, $\mu_{0}\Delta H =$ 5 T)
of the ErAl$_{2}$ powder evaluated using Eq. (\ref{eq:deltaS})
based on $\partial M/\partial T$-$\mu_{0} H$ data calculated from the $M$-$T$ curves
and the $M$-$\mu_{0}H$ curves (see the supplementary data), respectively.
Two $\Delta S_{M}$ curves almost agree with each other and peaks at $T_{\rm c}$.
This result indicates that the magnetic entropy change can be correctly evaluated
through the isofield magnetization measurements.
$\Delta S_{M}$ of the PIT wires were evaluated by the same procedure.}

\begin{figure}[t]
\centering
\includegraphics[width=70.00mm]{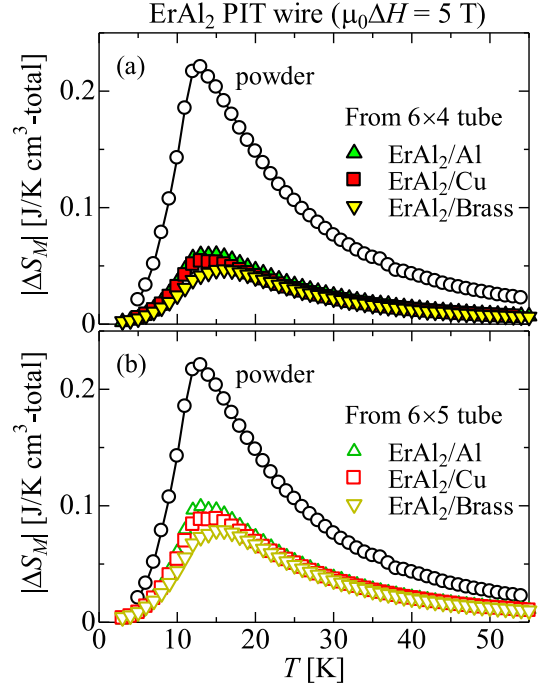}
\caption{(Color Online) Temperature dependence of $|\Delta S_{M}|$
for $\mu_{0} \Delta H =$ 5 T in various ErAl$_{2}$ PIT wires fabricated from
(a) 6$\times$4 and (b) 6$\times$5 tubes,
along with the data for the ErAl$_{2}$ powder.
For the PIT wires, $|\Delta S_{M}|$ represents the magnetic entropy change
of the wire with a total volume of 1 cm$^{3}$.}
\label{fig:deltaS}
\end{figure}

Figures \ref{fig:deltaS} (a) and \ref{fig:deltaS} (b) represent
the temperature dependence of $|\Delta S_{M}|$ for $\mu_{0} \Delta H =$ 5 T
per total volume of 1 cm$^{3}$ in various ErAl$_{2}$ PIT wires fabricated from
the 6$\times$4 and 6$\times$5 tubes.
The data for the ErAl$_{2}$ powder is also shown for the comparison.
The magnetic entropy change of the PIT wires exhibits
qualitatively similar characteristics as those of the powder sample,
while the magnitude is decreased by about 60-70\%.
This result is not surprising
because the volume fraction of ErAl$_{2}$ is reduced in the PIT wires.
In that sense, the data for the powder sample can be regarded as
$|\Delta S_{M}|$ of a hypothetical wire with 100\% ErAl$_{2}$ core material.
\textcolor{black}{
Indeed, $|\Delta S_{M}|$ becomes larger
in the case of the PIT wire fabricated from the 6$\times$5 tube,
namely, the larger filling rate of the core material.
Furthermore, when the filling rate is the same,
at the temperatures above 30 K, $|\Delta S_{M}|$ does not depend on the sheath material.}
These facts suggest that the volume fraction of the ErAl$_{2}$ core material mainly determines
the magnetic entropy change of the PIT wires.
On the other hand, we should notice the difference in $|\Delta S_{M}|$
between the PIT wires at around $T_{\rm c}$,
where the $|\Delta S_{M}|$ of the ErAl$_{2}$/Brass wire is significantly decreased.
A possible origin of which is discussed below.

\begin{figure}[t]
\centering
\includegraphics[width=70.00mm]{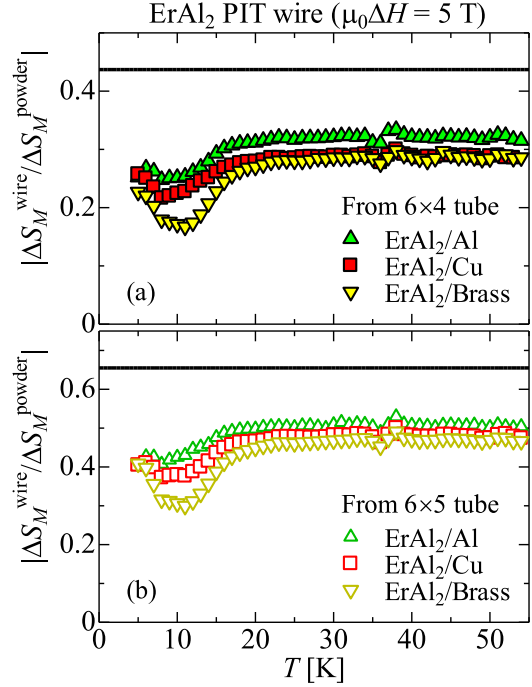}
\caption{(Color Online) Temperature dependence of
$|\Delta S_{M}^{\rm wire}/\Delta S_{M}^{\rm powder}|$
for the ErAl$_{2}$ PIT wires made from
(a) 6$\times$4 and (b) 6$\times$5 tubes (see the text).
Solid lines depict the theoretical volume fraction of the core material
expected from Fig. \ref{fig:SEM} with assuming no voids.}
\label{fig:VF}
\end{figure}

Here let us evaluate the ratios of the magnetic entropy change in each PIT wire
($|\Delta S_{M}^{\rm wire}|$) to that in the ErAl$_{2}$ powder ($|\Delta S_{M}^{\rm powder}|$),
which should correspond to the volume fraction of the core material.
Figures \ref{fig:VF}(a) and \ref{fig:VF}(b) show the temperature dependence of
$|\Delta S_{M}^{\rm wire}/\Delta S_{M}^{\rm powder}|$
calculated for the PIT wires made from  the 6$\times$4 and 6$\times$5 tubes.
One finds that the ratios take constant values at temperatures above 30 K.
This makes sense because the volume fraction should not change at any temperature.
On that account, we employ the mean value of the temperature-independent
$|\Delta S_{M}^{\rm wire}/\Delta S_{M}^{\rm powder}|$
as the actual volume fraction of ErAl$_{2}$ in the PIT wires,
being $\sim$ 0.30 for the wires made from the 6$\times$4 tube
and $\sim$ 0.49 for those made from the 6$\times$5 tube.
These values are about 70-75\% of the theoretical volume fraction
expected from the SEM images assuming no voids.
\textcolor{black}{
Funk et al. have reported that the volume fraction of La(Fe, Si, Co)$_{13}$ core is
about 85\% of the theoretical one,
even though pre-compacted raw materials were filled into a metal tube\cite{Funk-MTE-2018}.
In contrast, ErAl$_{2}$ powder was filled without any treatments in this study, 
implying that there can be more voids in our PIT wires compared with the La(Fe, Si, Co)$_{13}$ PIT wire.
Accordingly, the obtained values of the ErAl$_{2}$ core volume fraction seem to be reasonable.
With further decreasing temperature,
$|\Delta S_{M}^{\rm wire}/\Delta S_{M}^{\rm powder}|$ gradually decreases
and exhibits a dip structure near $T_{\rm c}$,
whose characteristic is noticeable with ErAl$_{2}$/Brass wires.
This behavior suggests that there is another contribution
that affects the magnetic entropy change of ErAl$_{2}$ itself in the PIT wires,
in addition to the decrease in the volume fraction of the core material.}

\begin{figure}[t]
\centering
\includegraphics[width=60.00mm]{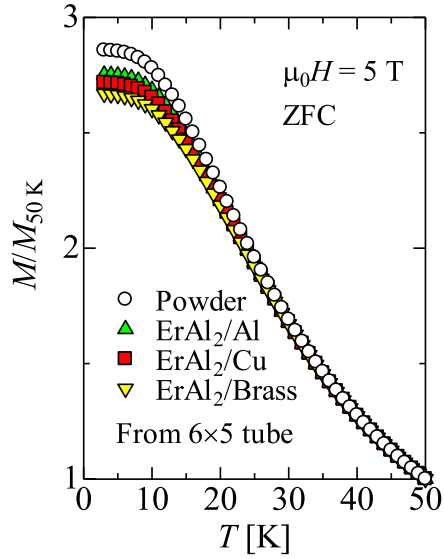}
\caption{(Color Online)
Temperature dependence of magnetization at 5 T
normalized by the magnetization at 50 K
for ErAl$_{2}$ powder and PIT wires made from the 6$\times$5 tube.}
\label{fig:M5T}
\end{figure}

Now we will discuss a possible origin of the extra reduction of 
$|\Delta S_{M}|$ around $T_{\rm c}$ observed in the PIT wires.
According to Eq. (\ref{eq:deltaS}), a decrease in $\Delta S_{M}$ results from
a decrease in ($\partial M/\partial T$)$_{H}$,
which occurs when $M$ decreases without changing the temperature dependence
and/or when the temperature dependence itself becomes more gradual.
To clarify this point, we plot $M/M_{\rm 50 \ K}$ at 5 T as a function of temperature in Fig. \ref{fig:M5T}
for the ErAl$_{2}$ powder and the PIT wires made from the 6$\times$5 tube.
The magnetizations show the same temperature dependence down to 30 K for all the samples,
but the rise in $M$ of the PIT wires is suppressed with decreasing temperature,
the trend is most significant in the ErAl$_{2}$/Brass wire.
This mild temperature variation does be the cause of
the decrease in ($\partial M/\partial T$)$_{H}$ for the PIT wires.
The difference in $M$-$T$ curves observed here resembles
those found in ferromagnetic materials with a uniaxial magnetic anisotropy
\cite{Zhang-PRL-2001,Luis-EPL-2006,Liu-PRM-2019},
in which ($\partial M/\partial T$)$_{H}$ becomes smaller
in the direction perpendicular to an easy axis of the magnetization.
Thus, the extra reduction of $|\Delta S_{M}|$ around $T_{\rm c}$ implies that
\textcolor{black}{
the PIT process induces a magnetic anisotropy in the ErAl$_{2}$ core material
with an easy axis perpendicular to the longitudinal direction of the wire.}

\textcolor{black}{
It is well known that a rolling process causes
a kind of magnetic anisotropy in magnetic materials
\cite{Chikazumi-JAP-1958,Chikazumi-JPSJ-1960,Chin-JAP-1967,Morita-JJAP-1979}.
This magnetic anisotropy is known to increase as the mechanical deformation increases,
and the latter usually increases with the stress on magnetic material during rolling.
On the other hand, in several studies on superconducting PIT wires
\cite{Grasso-APL-2001, Kumakura-PhysC-2002},
it has been pointed out that the higher the hardness of sheath material,
the stronger the stress on core material during cold working.
From these facts, the magnetic anisotropy induced by rolling is
expected to be large in the use of the harder tube in a PIT process.
In fact, since Vickers hardness is higher in the order of Al, Cu, and Brass,
the expectation is consistent with the result that
$|\Delta S_{M}|$ around $T_{\rm c}$ is most decreased
in the ErAl$_{2}$/Brass PIT wires.
Therefore, we conclude that the PIT process affects
the magnetocaloric properties of the ErAl$_{2}$ core material
through the induced uniaxial magnetic anisotropy.
However, the exact nature of the magnetic anisotropy
remains unclear at the present stage.
To get more insight, it is desirable to investigate the effect of annealing
that may control the plastic deformation.
}
\section{Conclusion}
We have fabricated the magnetocaloric wires of ErAl$_{2}$ cladded by
non-magnetic metal sheaths by using a powder-in-tube method combined with groove rolling.
These PIT wires exhibit magnetic entropy changes similar to that of the powder sample,
with their magnitude reduced due to the decrease in the volume fraction of the ErAl$_{2}$ core.
We propose that the PIT process affects the magnetocaloric properties of the core material
through \textcolor{black}{a kind of the induced uniaxial magnetic anisotropy}
and causes the extra reduction of the magnetic entropy change around $T_{\rm c}$.
There is still room for improvement of the magnetocaloric properties in the PIT wires
by annealing process and additional processes that increase the volume fraction of the core.
We believe that the wire-shaped magnetocaloric materials prepared by a PIT method would be
of benefit to the development of magnetic refrigerators for the hydrogen liquefaction.
\ack{
This work was supported by JST-Mirai Program Grant Number JPMJMI18A3, Japan.}

\end{document}